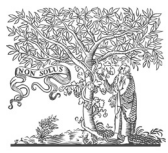
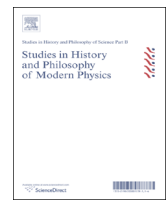
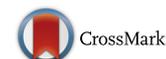

# Cosmology and convention

David Merritt

School of Physics and Astronomy, Rochester Institute of Technology, 54 Lomb Memorial Drive, Rochester, NY 14623, USA

ABSTRACT

I argue that some important elements of the current cosmological model are 'conventionalist' in the sense defined by Karl Popper. These elements include dark matter and dark energy; both are auxiliary hypotheses that were invoked in response to observations that falsified the standard model as it existed at the time. The use of conventionalist stratagems in response to unexpected observations implies that the field of cosmology is in a state of 'degenerating problemshift' in the language of Imre Lakatos. I show that the 'concordance' argument, often put forward by cosmologists in support of the current paradigm, is weaker than the convergence arguments that were made in the past in support of the atomic theory of matter or the quantization of energy.

© 2017 The Author. Published by Elsevier Ltd. This is an open access article under the CC BY-NC-ND license (http://creativecommons.org/licenses/by-nc-nd/4.0/).

## 1. Introduction

The idea that scientific theories contain 'conventional' aspects is attributed to Henri Poincaré (Poincaré, 1902). From his work on non-Euclidean geometries and higher spaces, Poincaré reached the conclusion that many elements of scientific theories which had been held to be fundamental truths were in fact just conventions. Thus any geometry can be adopted for space, if the necessary revision is made in the definition of a straight line. The laws of mechanics can likewise be interpreted as defining the concepts of force and inertial motion. While noting that some conventions might be more convenient than others, Poincaré asserted that any set of conventions could always be replaced by a different set without changing the content of a theory. In Poincaré's view, the parts of a scientific theory that are conventional cannot be said to be true or false; they are simply definitions, and as such, are immune to testing.

Unlike Poincaré, Pierre Duhem (1914) believed that experimental refutation of a theoretical system is possible. Nor did Duhem accept that any part of a theory could be singled out as definitional. Nevertheless, Duhem, like Poincaré, is often regarded as a conventionalist. Duhem noted that the prediction that a phenomenon will be observed is based on a set of premises, including laws, initial conditions, assumptions about the reliability of the experimental apparatus etc. In the face of a falsifying instance, the experimenter has no way of knowing which of these premises is false. Thus, no experiment or observation can ever be considered decisive against a particular hypothesis, and no hypothesis can be conclusively falsified.

Karl Popper was concerned with finding a criterion that demarcates science from non-science (or 'metaphysics'). He argued that falsifiability is such a criterion: scientific theories are falsifiable; non-falsifiable theories are non-scientific. Popper acknowledged the strength of the conventionalist position: "I regard conventionalism as a system which is self-contained and defensible" (Popper, 1959, p. 80). But Popper equated conventionalism with non-falsifiability, and he rejected it, in part because he saw conventionalism as impeding the growth of knowledge:

> Whenever the 'classical' system of the day is threatened by the results of new experiments which might be interpreted as falsifications according to my point of view, the system will appear unshaken to the conventionalist. He will explain away the inconsistencies which may have arisen … We, and those who share our attitude, will hope to make new discoveries; and we shall hope to be helped in this by a newly erected scientific system. Thus we shall take the greatest interest in the falsifying experiment. We shall hail it as a success, for it has opened up new vistas into a world of new experiences. (Popper, 1959, p. 80.).

Popper nevertheless acknowledged that the "conventionalist mode of thought" is useful, in that it can expose certain logical shortcomings in his doctrine of falsification:

I admit, a conventionalist might say, that the theoretical systems of the natural sciences are not verifiable, but I assert that they are not falsifiable either. For there is always the possibility of '… attaining, for any chosen axiomatic system, what is called its "correspondence with reality" ' (Carnap 1923, p. 100); and this can be done in a number of ways… Thus we may introduce *ad hoc* hypotheses. Or we may modify the so-called 'ostensive definitions'… Or we may adopt a sceptical attitude as to the reliability of the experimenter whose observations, which threaten our system, we may exclude from science on the ground that they are insufficiently supported, unscientific, or not objective, or even on the ground that the experimenter was a liar…. In the last resort we can always cast doubt on the acumen of the theoretician. (Popper, 1959, p. 81.).

Popper coined the term 'conventionalist stratagem' to describe these four ways of evading the consequences of a falsifying experiment. While admitting that there was no strictly logical basis on which to exclude such stratagems, he argued that in order to maintain falsifiability, conventionalist methods needed to be strictly avoided, and that "The only way to avoid conventionalism is by taking a *decision*: the decision not to apply its methods. We decide that, in the case of a threat to our system, we will not save it by any kind of *conventionalist stratagem*" (Popper, 1959, p. 82). Popper presented methodological rules for the practice of science that were designed to rule out the incorporation of conventionalist elements.[1]

Critics of Popper have debated whether falsification is a primary goal of scientists. Thomas Kuhn (1962) famously argued that the main occupation of scientists is not falsification but 'puzzle-solving,' an activity that implies uncritical acceptance of the current scientific paradigm. However, interpreted narrowly as a definition of the conventionalist program, Popper's list of stratagems need not be problematic. In what follows, I identify 'conventionalism' with Popper's list. A 'conventionalist' approach is defined as one which (whether deliberately or not) evades the consequences of a falsifying experiment or observation by the application of one or more of Popper's conventionalist stratagems.

## 2. The standard model of cosmology

At any given time, discrepancies exist between the predictions of an accepted scientific theory and the experiments or observations that test those predictions. Kuhn argued that most such discrepancies, which he called 'anomalies', are not viewed by scientists as falsifying instances. Rather, they are considered puzzles to be solved within the existing paradigm.

The standard model of cosmology[2] is not exceptional in this regard. The list of anomalies is impressively long, and some of them have persisted so stubbornly and for so long a time that they have achieved the status of 'named' problems. Examples include the 'Lithium problem' (Fields, 2011); the 'core-cusp problem' (de Blok, 2010); the 'missing satellites problem' (Moore et al., 1999); the 'too big to fail problem' (Boylan-Kolchin, Bullock & Kaplinghat, 2011); and the 'missing baryons problem' (McGaugh, 2008). In textbooks and review articles, these discrepancies are rarely described as falsifying; they are presented rather as problems that remain to be solved from within the existing paradigm.[3] Typical is the following statement by Malcolm Longair in the 2008 monograph *Galaxy Formation*: "There is no limit to the ingenuity of astronomers and astrophysicists in finding ways of reconciling theory and observation. As more parameters are included in the models, the easier it will be to effect the reconciliation of theory with observation" (p. 419).

At the same time, there *have* been instances since the 1960s where anomalies were interpreted by the community as being incompatible with the cosmological model as it existed at the time. A famous example is the discovery around 1998 that the expansion of the universe is accelerating, rather than decelerating as the standard model had predicted.

It is with the latter sort of discrepancy that this paper is concerned: that is: discrepancies that seem immune to reconciliation by (as Longair might say) adjusting the parameters of astrophysical theory. Three such instances are identified below. In each case, it will be argued that the response of the scientific community (whether intentionally or not) has been conventionalist in the sense defined by Popper. On this view, some essential components of the current, standard model of cosmology—including dark matter and dark energy—owe their existence to conventionalist stratagems.

## 3. Popper's "conventionalist stratagems"

Herbert Keuth (2005) provides a succinct re-statement of Popper's four conventionalist stratagems:

(i) we may introduce ad hoc hypotheses (which make refuting evidence seem irrelevant); (ii) we may modify the so-called ostensive definitions (so as to alter the content of a hypothesis and thus possibly its truth value); (iii) we may doubt the reliability of the experimenter (and declare his observations that threaten the tested theory to be irrelevant); (iv) we may doubt the acumen of the theoretician (who does not produce ideas that can save the tested theory). (Keuth, 2005, p. 72.).

Popper believed that scientists should avoid such stratagems, and to that end, he proposed a set of methodological rules that were designed to preserve falsifiability (Popper, 1959, chapter 4). Now, nothing in the present work is intended as *prescriptive*: neither the content of the current model of cosmology, nor the methodology that led to that content, are being critiqued here. Popper's prescriptivist rules are therefore not of direct interest; nor are the criticisms, by others, of those rules, of which there are many. However, in the process of specifying how falsifiability could be preserved, Popper sharpened and clarified the definitions of the four conventionalist stratagems, and those clarifications will be useful in what follows.

The first stratagem employs 'ad hoc hypotheses'.[4] Popper writes:

As regards *auxiliary hypotheses* we decide to lay down the rule that only those are acceptable whose introduction does not diminish the degree of falsifiability or testability of the system in question, but on the contrary, increases it…. If the degree of falsifiability is increased, then introducing the hypothesis has actually strengthened the theory: the system now rules out more than it did previously: it prohibits more. (Popper, 1959, pp. 82–83).

---

[1] In so doing, Popper showed himself to be a conventionalist with regard to methodology (Akinci, 2004).

[2] Here and below, the 'standard model of cosmology' refers to the 'ΛCDM [Lambda-cold-dark-matter] model' or the 'concordance' or 'benchmark' cosmological model as it is presented in current textbooks and review articles. That model purports to describe the universe going back to times as early as the era of 'big-bang nucleosynthesis' (BBN) and possibly earlier. The discussion in this paper refers to the evolution of the universe from the era of BBN until the present. There is nearly perfect unanimity concerning the predictions of the standard model over this interval of time; for a list of representative texts, see Table 1.

[3] An exception is Kroupa's (2012) closely-reasoned argument that many of the outstanding anomalies should be considered falsifying. See also Kroupa, Pawlowski, and Milgrom (2012).

[4] Popper uses the terms 'ad hoc hypothesis' and 'auxiliary hypothesis' synonymously.

By referring to "the system in question", Popper appears to be saying that the theory plus the auxiliary hypothesis should have a higher degree of falsifiability than the theory alone. Now, Ingvar Johansson (1975, p. 54) makes the point that an auxiliary hypothesis is only introduced when a theory appears to have been falsified. Suppose (as would normally be the case) that the auxiliary hypothesis explains the anomaly (in the way that 'dark energy' explains the observed acceleration of the universe, or 'dark matter' explains galaxy rotation curves). If the auxiliary hypothesis lacks potential falsifiers itself, the degree of falsifiability of the system (theory plus auxiliary hypothesis) would necessarily be reduced. In what follows, I will argue that the addition of an auxiliary hypothesis to a theory is conventionalist insofar as it lacks potential falsifiers itself.

The second conventionalist stratagem consists of changing the "ostensive definitions" in such a way that a previously falsifying observation is no longer inconsistent with the theory. Popper acknowledges that some changes in definitions are permissible, but "they must be regarded as modifications of the system, which thereafter has to be re-examined as if it were new" (Popper, 1959, pp. 83–84).

As regards the third and fourth stratagems, Popper says only "As to the two remaining points in our list…we shall adopt similar rules. Inter-subjectively testable experiments are either to be accepted, or to be rejected in the light of counter-experiments." (Popper, 1959, p. 84.) Following Freeman (1973), I equate Popper's 'intersubjective agreement' with 'objectivity', as that term is used by practicing scientists, and take his statement to refer to experiments or observations that have been confirmed by independent researchers. Now, stratagem number three consists of ignoring an observation or experiment for one reason or another. Here I make the (obvious) point that—if an observation or experiment made by one researcher is ignored by another researcher—it is quite possible that the latter will not give a *reason* for doing so. Indeed s/he may not mention the experiment at all; or if a reason is given, it may not be the actual reason. I therefore do not insist that the *reasons* for ignoring a falsifying experiment or observation be stated, but only that it *be* ignored, or at least that its falsifying consequences are ignored.

But which experiments or observations should be regarded as falsifying, so that in ignoring them, a scientist is engaging in conventionalism? Clearly some experimental results are more damaging to a theory than others. Popper notes that

> A theory is tested not merely by applying it, or by trying it out, but by applying it to very special cases—cases for which it yields results different from those we should have expected without that theory, or in the light of other theories. In other words we try to select for our tests those crucial cases in which we should expect the theory to fail if it is not true. (Popper, 1962, p.112).

Further on, Popper defines a test as "severe" if it confronts a prediction that was "highly improbable in the light of our previous knowledge" (Ibid, p. 220). I will decide that a decision to ignore a discrepancy is conventionalist, to the degree that the ignored observation constitutes a "severe" test in the sense defined by Popper.

Popper's fourth stratagem consists of "cast[ing] doubt on the acumen of the theoretician" when s/he fails to find a way to explain the observation within the current theoretical system. As noted above, theoretical astrophysicists devote considerable effort to reconciling observational anomalies with the standard model of cosmology; such activity constitutes part of what Kuhn calls 'normal science'. Popper's fourth conventionalist stratagem is essentially identical to Kuhn's puzzle-solving, and I will take it for granted in what follows.[5] The focus here is on discrepancies that seem immune to reconciliation by 'adjusting the parameters' of astrophysical theory.

## 4. Dark matter

Newton's laws of motion relate acceleration to force; his law of universal gravitation relates gravitational force to mass.[6] Astrophysical observations provide an opportunity to test Newton's laws in a regime that is not accessible in the laboratory, or indeed anywhere in the Solar System: the regime of very low acceleration.[7] Stars or gas clouds orbiting near the outskirts of galaxies like the Milky Way are used as 'test particles'; a measured velocity, $V$, is converted into a (centripetal) acceleration, $a$, via $a = V^2/r$ where $r$ is the distance of the star or gas cloud from the center of the galaxy, under the assumption that motion takes place in a circular orbit. The measured acceleration is then compared with the acceleration predicted by Newton's law of gravitation. The two results are found to be in systematic and spectacular disagreement (Rubin, Thonnard & Ford, 1978; Bosma, 1981). Measured rotation velocities are greater, and sometimes much greater, than would be expected if the force responsible for the acceleration was produced by the observed mass (stars, gas) of the galaxy. Empirically, the discrepancy manifests itself wherever in a galaxy the acceleration predicted by Newton's law of gravitation falls below a value $a_0 \approx 10^{-10}$ m s$^{-2}$ (Milgrom, 1983; Sanders, 1990; Begeman, Broeils & Sanders, 1991; McGaugh, 2011). In the Milky Way, this occurs at distances from the galaxy center greater than about 6 kpc[8] (Sellwood & Sanders, 1988)[9]; by comparison, the Sun's distance from the Galactic center is about 8 kpc.

One could imagine responding to this anomaly by modifying Newton's laws of motion or his law of gravitation or both (Milgrom, 1983). Such modifications would be falsifiable, in the sense of making definite predictions about $V$ as a function of $r$ in a galaxy with a known distribution of mass.

The standard model of cosmology deals with this anomaly in a different way: via an auxiliary hypothesis. It is postulated that galaxies are embedded in 'dark matter haloes': approximately spherical systems composed of some kind of distributed matter that does not interact significantly with radiation but which does generate and respond to gravitational forces. Newton's laws of motion are retained; velocities predicted from those laws are larger because the total mass (normal plus dark) that generates the gravitational force is larger. In order to be consistent with the run of $V$ vs. $r$ in observed galaxies, the density of the dark matter would need to vary, approximately, as the inverse square of distance from the galaxy's center, and the normalization of the dark matter density (its value at some specified $r$) would need to be adjusted so that the enclosed, total mass begins to be dominated by dark matter at radii where the acceleration due to the normal matter falls below $\sim a_0$. Under this hypothesis, the total amount of dark matter in a galaxy would typically exceed the total amount of ordinary matter (stars, gas) by an order of magnitude or more.

---

[5] But see note 3.
[6] The descriptions of gravitation and motion in the standard model of cosmology are Einstein's, not Newton's, but the predictions of these two theories are indistinguishable in the low-acceleration regime considered in this section.
[7] This follows the sometimes confusing practice of astrophysicists to use the term 'acceleration' both to mean 'rate of change of velocity' and 'gravitational force per unit of mass' (where 'mass' refers to the body on which the force acts).
[8] kpc = kiloparsec = $3.09 \times 10^{21}$ cm.
[9] This statement assumes that there is no density inhomogeneity that increases the local acceleration above the mean value determined by the overall mass of the Galaxy. In the Solar system, for instance, the acceleration exceeds $a_0$ everywhere due to the presence of the Sun.

When attempting to reproduce the rotation curve (i.e. the plot of *V vs. r*) of an individual galaxy, the parameters describing the putative dark-matter halo are typically varied, arbitrarily, in order to give the best fit to the data. In this limited sense, the dark matter hypothesis can be said to be non-falsifiable, since essentially any observed rotation curve can be fit by adjusting the assumed dark matter density appropriately.

This aspect of non-falsifiability of the dark matter hypothesis need not be considered fundamental. Following Duhem, one can argue that the dark matter distribution around a given galaxy is determined by that galaxy's initial conditions and its unique history of evolution and interaction with other galaxies. Since these additional factors can not be well known, the best one might hope for theoretically is to predict certain statistical or average properties of dark matter: the average ratio of dark to luminous matter; the relative degree of central concentration of the two components; the degree of clustering on super-galactic scales etc.

Attempts starting in the 1980s to do this quickly revealed the need to augment the dark-matter hypothesis in one important way: the dark matter must be 'cold', i.e., the particles[10] making up the dark matter must have been moving non-relativistically at the time of structure formation; otherwise they would move so fast as to preclude their clumping gravitationally into structures with the sizes and densities inferred for the dark haloes. This condition on the velocities of dark matter particles can be converted into a condition on the particle mass, $m_\chi$, under certain, additional assumptions about how the dark matter 'decoupled' from ordinary matter in the early universe. In order for the dark matter particles to be sufficiently 'cold' (i.e. slowly moving), these additional assumptions require $m_\chi \gtrsim 1$ GeV/$c^2$.[11] Numerical simulations of the gravitational clumping of 'cold dark matter' have been shown to reproduce certain statistical properties of the distribution of galaxies and galactic systems, although many anomalies (in the sense defined by Kuhn) remain; indeed several of the 'named' problems mentioned in Section 2 refer to a particular manner in which these simulations fail to reproduce the observations of galaxies and galactic systems.

What would constitute a crucial experiment for testing the cold dark matter hypothesis? — Clearly, an experiment that is sensitive to the presence or absence of the dark matter particles themselves.[12] Now, a statement like "Dark matter particles exist" is an existential generalization, hence it is verifiable but not falsifiable. (A single dark-matter particle could always exist, undetected, in a galaxy far away.) However, as Popper points out (Popper, 1983, pp. 178–185), existential statements are often accompanied by qualifications arising out of the scientific context in which they are presented, rendering them falsifiable. In the case of cold dark matter, the particles that are postulated to make it up must have a particular local density and velocity distribution if they are to explain the Milky Way rotation curve. Some of the hypothesized particles must therefore be passing, at any given moment, through an Earth-based laboratory, with a distribution of velocities that is at least approximately known; an experimental demonstration of their absence would constitute a falsification. Now, the Milky Way rotation curve implies a mass density of the dark matter of about 0.4 GeV/$c^2$ cm$^{-3}$ at the location of the Earth (Catena and Ullio, 2012; Pato, Iocco & Bertone, 2015). The predicted *number* density of the dark matter particles is therefore $\sim 0.04/m_{10}$ cm$^{-3}$ where $m_{10} = m_\chi/(10 \text{ GeV}/c^2)$ is the assumed particle mass in units of 10 GeV/$c^2$. Elastic collisions between a dark matter particle and normal matter could transfer kinetic energy to the latter; passage of a dark matter particle through a laboratory detector could in principle be observed using one of a number of well-established techniques based on ionization, scintillation or calorimetry. These ideas are the basis for so-called 'direct detection' experiments, a number of which have been in operation since the mid 1980s.[13]

These experiments have so far failed to detect dark matter particles.[14] But regardless of how sensitive the experiments become, non-detection will never constitute a falsification of the cold-dark-matter hypothesis. Laboratory detection of a dark matter particle requires that both the number density of the particles at the Earth, and the cross-section, $\sigma_\chi$, for a dark matter particle to interact with normal matter, be sufficiently large. As discussed above, the number density depends on the particle mass $m_\chi$, for which a plausible lower limit can be placed. But nothing whatsoever is known about $\sigma_\chi$. A non-detection in the laboratory therefore implies only an upper limit on $\sigma_\chi$ at any given value of $m_\chi$. If no detection occurs, it may simply mean that the dark matter particles have a very low cross-section for interaction with normal matter; it can not be interpreted to mean that the particles are not present in the detector.[15]

No particle having the properties of the putative dark matter particle exists within the standard model of particle physics, and theoretical speculations about the probability of direct detection are often presented in the context of some alternative to, or extension of, that model. Probably the most common choice is supersymmetry.[16] At the time of the first direct-detection experiments, an interaction cross section $\sigma_\chi$ of approximately $10^{-39}$ cm$^2$ (at $m_\chi = 50$ GeV) was considered 'natural' by theoretical physicists working on supersymmetry. Such values for $\sigma_\chi$ were ruled out already by the mid 1990s (Fig. 1, left panel). Current experimental upper limits on $\sigma_\chi$ are about six orders of magnitude smaller; particle physicists have accommodated the persistent non-detection by constructing supersymmetric models in which the particles constituting the dark matter have correspondingly smaller interaction cross sections (Fig. 1, right panel). In principle, adoption of a well-defined particle physics model could render the cold-dark-matter hypothesis falsifiable, by allowing definite predictions to be made about the properties of the dark matter particles. At least at the present time, theoretical alternatives to the standard model of particle physics are not well enough defined to do that.[17]

---

[10] That the dark matter is particulate is generally assumed, although other possibilities have been discussed; see Bertone (2010, Chapters 9–11).

[11] GeV=billion electron volts, c=speed of light. One GeV/$c^2$ is a mass of $1.78 \times 10^{-27}$ kg. The mass of the proton is 0.94 GeV/$c^2$.

[12] Dark matter particles might also be created in particle accelerators, like the Large Hadron Collider. In the words of G. Bertone and J. Silk, "The discovery of new particles [in accelerator experiments] would be per se an extraordinary discovery, but it might not allow a straightforward identification with a successful DM [dark matter] candidate... progress will only be achieved by observing cold dark matter directly" (Bertone, 2010, p. 10). Similar considerations apply to so-called 'indirect detection' methods (as recently reviewed by Gaskins, 2016), which require a number of additional assumptions before an observed event can be interpreted in terms of dark matter.

[13] For recent reviews of dark-matter direct detection techniques and results see Cerdeño & Green (2010), Cushman et al. (2013), Olive et al. (2014), Marrodán Undagoitia & Rauch (2016).

[14] One group has claimed detection (Bernabei et al., 2000, 2013). This claim is not widely credited, primarily because other experiments, some even more sensitive, have failed to verify it (see review by Marrodán Undagoitia & Rauch, 2016).

[15] Less plausible hypotheses could also be invented to explain the lack of a detection; for instance, that the Earth is located in a local "hole" with respect to the dark matter.

[16] See e.g. Olive et al. (2014) for a review of the motivation behind supersymmetry. In supersymmetric models, every known elementary particle has a supersymmetric partner, with a spin differing by half a unit.

[17] One might argue, rather, that the direct detection experiments have the potential to falsify some proposed extensions to the standard model of particle physics, at least in the context of the standard cosmological model.

## 5. Dark energy

That the universe is expanding follows from the observed fact that light from distant galaxies is redshifted, i.e., features of known rest-wavelength in the spectra of galaxies are shifted toward longer wavelengths, indicating motion away from us. Furthermore the speed of recession (computed via Doppler's formula) is observed to increase with distance to the galaxy, in roughly a linear fashion ('Hubble's law'). Now, in the absence of forces other than gravity, Einstein's equations predict that the rate of expansion of a homogeneous and isotropic universe should decrease over time, since the gravitational attraction between all the matter in the universe continually opposes the expansion. Depending on the amount of matter in the universe, the universe would either be expected to expand without limit, although at an ever decreasing rate ('open Universe'), or it could reach a maximum size and then recollapse under the influence of gravity ('closed Universe'). The third possibility, a 'critical Universe', would tend asymptotically to a state of zero expansion after infinite time and after reaching infinite extent.

Which of these three possibilities describes the actual universe can be determined from observations, at least if Einstein's equations are assumed to be correct. One technique is to extend the 'Hubble diagram', the plot of Doppler shift *vs.* distance, to very distant galaxies, distant enough that (roughly speaking) their light was emitted at a time when the cosmological rate of expansion was significantly different than it is now. Starting around 1997, attempts to do this reached the conclusion that the expansion is *accelerating*, not decelerating as had been predicted (Riess et al., 1998; Perlmutter et al., 1999). Furthermore the data suggest that the Universe has been in its accelerating phase for roughly the last six billion years.

One could respond to this anomaly by modifying Einstein's theory of gravity, in such a way that an accelerating universe would be predicted in spite of the tendency of gravity to decelerate the expansion (Dvali, Gabadadze & Porrati, 2000; Freese and Lewis, 2002; Carroll, Duvvuri, Trodden & Turner, 2004). Alternatively, one could retain Einstein's equations, but relinquish the assumption of homogeneity. If the Earth lies near the center of a large ($\sim 10^3$ Mpc[18]) region of underdensity (compared with the large-scale average), the Hubble diagram, as measured from the Earth, could have the observed form, even in the absence of a cosmic acceleration (Garcia-Bellido and Haugbølle, 2008). Both hypotheses are falsifiable: in the former case via laboratory experiments (Gubser and Khoury, 2004) or tests of gravity on solar-system scales (Dvali 2003); in the latter case, by galaxy counts or observations of asymmetries in the big-bang relic radiation (Karachentsev, 2012; Caldwell and Stebbins, 2008)[19].

The standard model of cosmology deals with this anomaly in a different way: via an auxiliary hypothesis. It is postulated that the universe is filled with a fluid, called 'dark energy', that has whatever properties are needed to convert the predicted cosmological deceleration into an acceleration, and in just such a manner as to reproduce the observed dependence of galaxy redshift on distance. Einstein's theory of gravity is retained, as are the other assumptions (large scale homogeneity and isotropy) that underlie the standard cosmological model. Mathematically, dark energy appears as a new set of terms in the stress-energy tensor $T_{\mu\nu}$ of Einstein's field equations; the properties of the dark energy that must be specified in order to solve that equation are its energy density $\varepsilon$ and its pressure $p$, and their dependence on the cosmological scale factor. In order to reproduce the observed acceleration (in spite of the tendency of gravity to decelerate the expansion), dark energy must have one surprising property: its pressure must be negative, i.e. $p < 0$.[20]

Mathematically, there is a choice for $\varepsilon$ and $p$ that is particularly convenient: the energy density is set to a constant value (with respect to time), and the pressure is also assumed to be constant and to equal $-\varepsilon$. With these additional *Ansätze*, Einstein's equation,

$$G_{\mu\nu} = 8\pi G T_{\mu\nu},$$

becomes simply[21]

$$G_{\mu\nu} = 8\pi G T_{\mu\nu} - \Lambda g_{\mu\nu} \tag{1}$$

where $T_{\mu\nu}$ is the energy–momentum tensor of the matter (normal plus dark) alone and $\Lambda$ is a constant given by

$$\Lambda = 8\pi G \varepsilon / c^2 \tag{2}$$

with $c$ the speed of light. In this (standard) version of the dark-energy hypothesis, only a single new parameter, $\Lambda$, appears in the equations that describe the evolution with time of the cosmic scale factor.[22] Given this additional parameter to vary, the observed Hubble diagram can be fit; the data do not yield a unique value of $\Lambda$, rather they yield an approximately linear relation between $\Lambda$ and the assumed density of matter (normal plus dark): the larger the first, the larger the second. Given likely values of the matter density (derived using the sort of data discussed in the previous section), one infers a value of $\Lambda$ such that the equivalent density of the dark energy, $\rho_\Lambda \equiv \varepsilon/c^2 = \Lambda/8\pi G$, is of the same order as the density of matter at the current epoch.[23]

The dark energy hypothesis allows one to fit any observed cosmic expansion history by adjusting the dependence of $\varepsilon$ and $p$ on time[24] (Woodard, 2007). In this limited respect, the dark energy hypothesis is not falsifiable. The situation here is similar to that regarding dark matter, in the sense that the density of dark matter around any galaxy can always be adjusted in such a way as to reproduce any observed rotation curve.

In the case of dark matter, it was argued that a critical experiment would be one that was sensitive to the presence of the dark matter particles. Can one imagine designing a similar experiment that tests the dark energy hypothesis?

The straightforward answer is "no". Whereas the particulate nature of dark matter is taken nearly for granted by adherents to the ΛCDM model,[25] dark energy is sufficiently mysterious that nothing even approaching a consensus exists regarding its fundamental nature. The vagueness of the dark-energy hypothesis mitigates against its testability, to an even greater degree than in the case of dark matter.

---

[18] Mpc=megaparsec=$3.09 \times 10^{24}$ cm.

[19] In fact, there is evidence for a substantial underdensity on a scale of 300 Mpc around the Sun (Keenan, Barger & Cowie, 2013).

[20] One is reminded of the conclusion reached by some chemists in the 18th century that phlogiston must have a negative mass, in order to reproduce the observed gain in weight of some metals after burning; see Partington and McKie (1981).

[21] $g_{\mu\nu}$ is the 'metric tensor' and $G_{\mu\nu}$ is the 'Einstein tensor', both of which are related algebraically to space–time curvature.

[22] This is the $\Lambda$ that lends its name to ΛCDM.

[23] Since the matter density decreases with time as the universe expands, this near coincidence implies that we are living at a special time. This is one of several ways in which the standard model appears to be uncomfortably 'fine-tuned' (see e.g. Schneider, 2015, pp. 204–209).

[24] Cosmologists typically express this freedom in terms of a dependence of $\varepsilon$ on $p$, the so-called 'equation of state' of the dark energy, rather than a dependence of either quantity on time.

[25] In the words of Funk (2015): "Today, it is widely accepted that dark matter exists and that it is very likely composed of elementary particles." The technically challenging (and very expensive) direct-detection experiments described in the previous section would hardly be carried out (or funded) if this were not the case.

The falsifiability of the dark energy hypothesis might be improved by specifying more clearly the nature of the dark energy. But this is not assured. There is a wide class of dark-energy models which can be shown to have negligible experimental consequences on scales less than $\sim 10^5$ pc (Frampton, 2004, 2006; Branchina, Liberto & Lodato, 2009).

## 6. The mass discrepancy–acceleration relation

The *prima facie* case for dark matter is based on the fact that velocities observed in the outskirts of galaxies are higher than predicted by Newton's laws: that is: $V > V_{\text{Newton}}$. This discrepancy is striking enough, but there is an even more remarkable property of the rotation curve data (Milgrom, 1988; Sanders, 1990; Sancisi 2004; McGaugh, 2004). It turns out that the *degree* to which the measured velocities are too high (compared with the Newtonian prediction) is strictly predictable, point by point, in every galaxy so far observed, given only the local density of *normal* matter. In other words: the dark matter—which, putatively, dominates the galaxy's total mass budget and determines $V$—somehow knows to distribute itself in a way that strictly respects the distribution of the *normal* matter in every galaxy.

One aspect of this empirical relation was mentioned above: observed velocities exceed predicted velocities only where the gravitational acceleration[26] produced by the normal matter, $a_{\text{Newton}}$, falls below a certain, apparently universal, value $a_0$ (Milgrom, 1983; Sanders, 1990). This is remarkable enough, since nothing in Newton's or Einstein's descriptions of gravity or motion predicts that the behavior of a dynamical system should change, qualitatively, below some particular acceleration.[27]

But what is even more mysterious is that the local ratio between observed and expected velocities, $V/V_{\text{Newton}}$, is found to be *strictly predictable* given only the observed distribution of *normal* matter[28] in the galaxy. This result is illustrated in Fig. 2, which plots the 'mass discrepancy–acceleration' relation — so called because $V/V_{\text{Newton}}$ measures the 'missing mass', i.e. the degree to which the gravitational acceleration exceeds that which would be produced by the observed matter. The relation is empirical, but it is very tight (as empirical relations in astrophysics go); the scatter appears to be attributable entirely to measurement errors, i.e., the *intrinsic* scatter is consistent with zero (McGaugh, 2004).

At points in a galaxy where $a_{\text{Newton}} >> a_0$, one observes $V \approx V_{\text{Newton}}$; Newton's laws apparently hold and there is no need for dark matter. This is the case near the centers of some galaxies, e.g., the Milky Way. At the other extreme, where $a_{\text{Newton}} << a_0$, which is typically the case in the outskirts of galaxies, the empirical relation of Fig. 2 is well approximated as

$$V/V_{\text{Newton}} \simeq (a_0/a_{\text{Newton}})^{1/4} \quad \text{when} \quad a_{\text{Newton}} << a_0 \qquad (3)$$

If one interprets the left hand side of Eq. (3) as a measure of the dark matter (which is the interpretation given to it in the standard model), the right hand side implies that the dark matter distribution is strictly tied to that of the normal matter, and furthermore that the relation is a simple algebraic one.[29]

There is a class of galaxies, the so-called low-surface-brightness galaxies, in which the density of normal matter is so low that $a_{\text{Newton}} << a_0$ everywhere, even near the center (Bothun, Impey & McGaugh, 1997). In the standard model, such galaxies would be 'dark-matter dominated' everywhere: the gravitational force throughout the galaxy would be due almost entirely to the dark matter. The standard model has produced no algorithm for predicting, in detail, the distribution of dark matter in such a galaxy (or any particular galaxy for that matter), and there is no reason to expect the dark matter to be rigorously controlled by the much smaller amount of normal matter. Nevertheless, low surface brightness galaxies accurately obey Eq. (3) (McGaugh & de Blok, 1998; de Blok & McGaugh, 1998).

Explaining observed correlations is an aspect of what Kuhn calls 'normal science'. What distinguishes the mass discrepancy–acceleration relation is that it seems immune to such explanation, at least from within the paradigm of the standard cosmological model, no matter (as Longair might say) how many astrophysical parameters are varied (Wu and Kroupa, 2015). Even if a relation like that of Eq. (3) were established at some early time between the dark and normal matter (no mechanism for doing this has yet been proposed), it could not persist in the face of interactions (tidal encounters, mergers) between galaxies, or processes like star formation and galactic winds, which would affect the dark and normal matter in substantially different ways, and the effects of which would vary, presumably randomly, from galaxy to galaxy.[30] But the correlation plotted in Fig. 2 shows no evidence of stochasticity, contingency, or dependence on environment. It has more the manifestation of a natural law (McGaugh, 2014).

Framed as a prediction, the mass discrepancy–acceleration relation clearly satisfies Popper's criterion for a 'severe' test ("highly improbable in the light of our previous knowledge"). No attempt has been made to account for this anomaly within the framework of the standard cosmological model. Instead, the mass discrepancy–acceleration relation has been dealt with via the third of Popper's conventionalist stratagems: It has been ignored.

Simulations of galaxy formation carried out within the standard model—examples include Brook et al. (2012), Vogelsberger et al. (2014), and Wellons et al. (2015)—routinely include comprehensive comparisons of the properties of the simulated galaxies with observed properties and correlations. By 'adjusting the parameters' of star formation, gas dynamics, and other 'sub-grid' physics, these simulation studies sometimes do a reasonable job of reproducing the observed relations. But comparisons with the mass discrepancy–acceleration relation (and citations to the discovery papers) are lacking from these and similar studies.

The same is true with regard to textbooks on galaxy formation and cosmology. Before documenting that statement, it is necessary to determine on what date the relation was first established and published. The fact that galaxy rotation curves 'know' about the distribution of light (even where the dark matter, supposedly, dominates) has been known at least since 1980 (Rubin, Burstein & Thonnard, 1980). The existence of a universal acceleration scale, $a_0$, below which observed disk-galaxy rotation velocities exceed $V_{\text{Newton}}$, was well established by 1983 (Milgrom, 1983). Correlation plots like the one shown here in Fig. 2 were published at least as early as 1990 (Sanders, 1990; McGaugh, 1998); the relation appears to have been given its current name in 2004 (McGaugh, 2004).

The role of textbooks in defining a scientific paradigm has been discussed by Thomas Kuhn:

> They [the textbooks] address themselves to an already

---

[26] See footnote 7. In this paragraph, 'acceleration' means 'force per unit mass'.
[27] On the other hand, until the measurement of galactic rotation curves in the 1970s, these theories were never tested in regimes of such low acceleration; $a_{\text{Newton}} >> a_0$ everywhere in the Solar System.
[28] The directly observable quantity is not mass but radiation: the light (from the stars) and the radio emission (from the gas). Techniques exist for relating those observables to mass with relatively little uncertainty (e.g. Lelli, McGaugh & Schombert, 2016).
[29] The reader may wonder how it ever occurred to anyone to plot the rotation-curve data in this non-intuitive way, rather than as (for instance) $V/V_{\text{Newton}}$ vs radius. The intriguing answer is given in Famaey & McGaugh (2012).
[30] This randomness is clearly observed in the distributions of stars and gas.

**Table 1**
Graduate-level texts on cosmology and/or galaxy formation published after 2004. Q1="Is the failure of laboratory experiments to detect dark matter discussed?" Q2="Is the existence of a universal acceleration scale $a_0$ discussed?" Q3="Is the mass discrepancy–acceleration relation discussed?".

| Date | Author(s) | Title | Q1 | Q2 | Q3 |
| --- | --- | --- | --- | --- | --- |
| 2005 | J.A. Gonzalo | Inflationary Cosmology Revisited: An Overview of Contemporary Scientific Cosmology After the Inflationary Proposal | No | No | No |
| 2005 | J.F. Hawley, K.A. Holcomb | Foundations of Modern Cosmology | Yes | No | No |
| 2005 | F. Hoyle, G. Burbidge, J.V. Narlikar | A Different Approach to Cosmology | No | No | No |
| 2005 | D.-E. Liebscher | Cosmology | No | No | No |
| 2005 | V. Mukhanov | Physical Foundations of Cosmology | No | No | No |
| 2005 | S. Phillipps | The Structure and Evolution of Galaxies | No | No | No |
| 2007 | L.S. Sparke, J.S. Gallagher | Galaxies in the Universe: An Introduction | No | No | No |
| 2007 | W.C. Keel | The Road to Galaxy Formation | No | No | No |
| 2008 | S. Weinberg | Cosmology | Yes | No | No |
| 2008 | M. Longair | Galaxy Formation | Yes | No | No |
| 2008 | J.M. Overduin, P.S. Wesson | The Light/Dark Universe: Light from Galaxies, Dark Matter and Dark Energy | Yes | No | No |
| 2008 | M. Giovannini | A Primer on the Physics of the Cosmic Microwave Background | No | No | No |
| 2009 | A. Liddle, J. Loveday | The Oxford Companion to Cosmology | No | Yes | No |
| 2010 | L. Amendola, S. Tsujikawa | Dark Energy: Theory and Observations | No | No | No |
| 2010 | S. Serjeant | Observational Cosmology | Yes | No | No |
| 2010 | H. Mo, F. van den Bosch, S. White | Galaxy Formation and Evolution | No | No | No |
| 2010 | T.-P. Cheng | Relativity, Gravitation and Cosmology: A Basic Introduction | No | No | No |
| 2010 | J. Rich | Fundamentals of Cosmology | Yes | No | No |
| 2010 | P. Ruiz-Lapuente | Dark Energy: Observational and Theoretical Approaches | No | No | No |
| 2010 | Y. Wang | Dark Energy | No | No | No |
| 2011 | S. Capozziello and V. Faraoni. | Beyond Einstein Gravity: A Survey of Gravitational Theories for Cosmology and Astrophysics | No | No | No |
| 2011 | S. Matarrese et al. | Dark Matter and Dark Energy: A Challenge for Modern Cosmology | Yes | No | No |
| 2012 | M. Bojowald | The Universe: A View from Classical and Quantum Gravity | No | No | No |
| 2012 | G. F. R. Ellis et al. | Relativistic Cosmology | No | No | No |
| 2013 | P. Peter, J.-P. Ulan | Primordial Cosmology | Yes | Yes | No |
| 2014 | J. Einasto | Dark Matter and Cosmic Web Story | No | No | No |
| 2015 | T.-P. Cheng | A College Course on Relativity, Gravitation and Cosmology: A Basic Introduction | Yes | No | No |
| 2015 | C. Deffayet et al. | Post-Planck Cosmology | No | No | No |
| 2015 | M. H. Jones et al. | An Introduction to Galaxies and Cosmology | Yes | No | No |
| 2015 | A. Liddle | An Introduction to Modern Cosmology | Yes | No | No |
| 2015 | D. Majumdar | Dark Matter: An Introduction | Yes | No | No |
| 2015 | M. Roos | Introduction to Cosmology | Yes | No | No |
| 2015 | P. Schneider | Extragalactic Astronomy and Cosmology: An Introduction | Yes | No | No |
| 2016 | D. Lyth | Cosmology for Physicists | No | No | No |

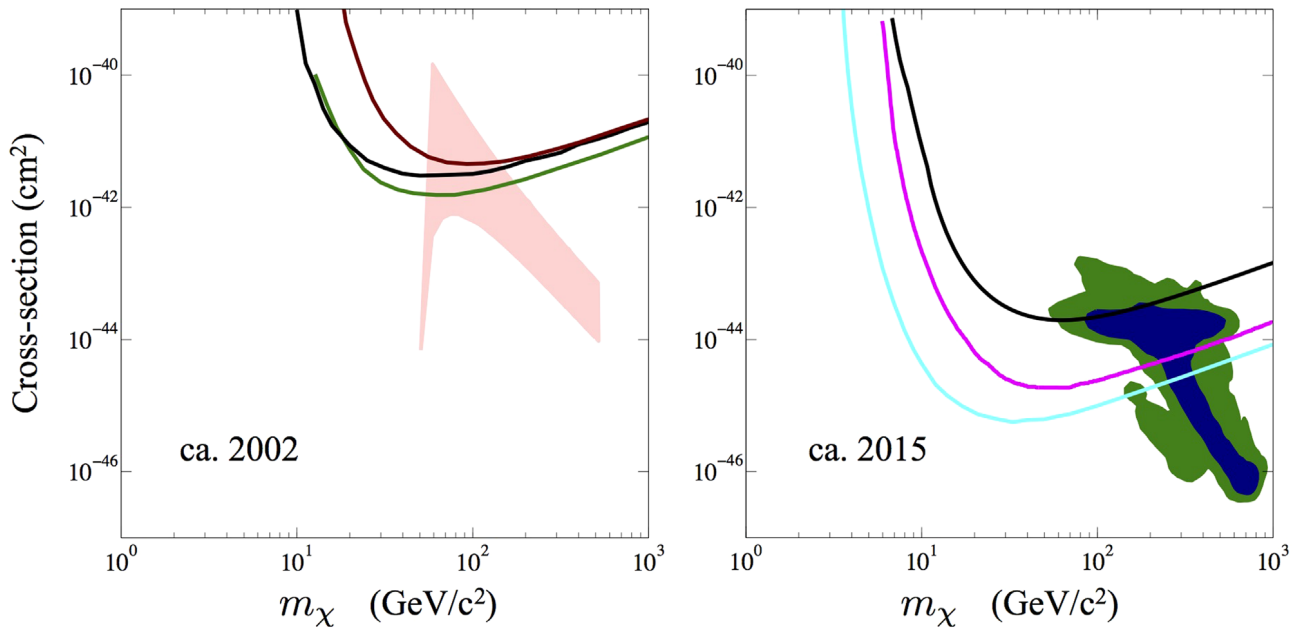

**Fig. 1.** Results of dark matter direct-detection experiments. The abscissa is the mass of the putative dark matter particle; the proton has a mass of about one in these units. The ordinate is the putative cross section for interaction of a dark matter particle with nuclei of normal matter. The curves are upper limits set by various experiments; regions that lie above these curves are excluded with high confidence. The left panel shows results from experiments that had been completed by about 2002 (*red*: Benoit et al., 2001, EDELWEISS; *black*: Abusaidi et al., 2000, CDMS; *green*: Barton et al., 2003, ZEPLIN). The right panel shows experiments that had been completed by about 2015 (*black*: Agnese et al. 2015, CDMS II; *magenta*: Aprile, 2012, XENON100; *cyan*: Akerib et al., 2016, LUX). The shaded regions represent theoretical expectations, based on some hypothesized extension of the standard model of particle physics. The light red region in the left panel is from Ellis, Ferstl, and Olive (2000); the blue and green regions (68% and 90% confidence respectively) are from Trotta et al. (2008). As the experimental limits have become tighter, cosmologists have postulated ever more extreme properties for the dark matter particles. (For interpretation of the references to color in this figure legend, the reader is referred to the web version of this article.)

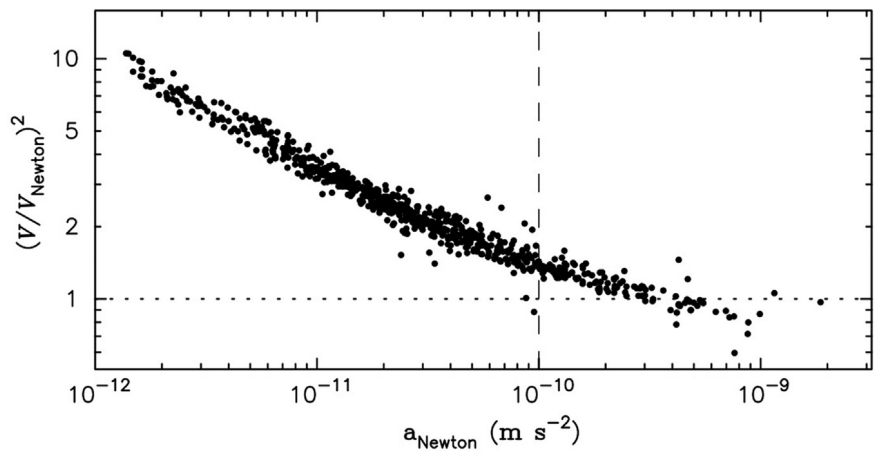

**Fig. 2.** The 'mass discrepancy–acceleration relation' (Sanders, 1990; McGaugh, 1998, 2004). Each point is a measured value, at one position in one galaxy. Points toward the left (low acceleration) are farthest from the galaxy center. Many different galaxies and galaxy types (size, morphology, surface brightness) are included. The fact $V \geq V_{\text{Newton}}$ is the basis for the dark matter hypothesis. This figure shows in addition that the observed $V$ is strictly predictable, at every location in every galaxy, given only the value of the gravitational force from the normal matter. Such a relation is extremely difficult to understand under the dark matter hypothesis, for the reasons discussed in the text. The dashed vertical line is roughly $a_0$; wherever in a galaxy $a_{\text{Newton}} >> a_0$, the observed $V$ is correctly predicted by Newton's laws, i.e. $V \approx V_{\text{Newton}}$.

articulated body of problems, data, and theory, most often to the particular set of paradigms to which the scientific community is committed at the time they are written. Textbooks themselves aim to communicate the vocabulary and syntax of a contemporary scientific language. (Kuhn, 1962, p. 136.).

If Kuhn's characterization of textbooks is correct, they are an excellent source of information about the current state of a scientific theory.

Here I take the conservative view that only textbooks published in 2005 or later could be expected to discuss either the existence of a universal acceleration scale, or the mass discrepancy–acceleration relation. (Included in that subset are textbooks that were first published prior to 2005 but which appeared in revised or updated editions after 2004.)[31] A list of such texts is given in Table 1. The list is intended to be complete: it includes every, or nearly every, book published during this period that presents the subjects of cosmology and/or galaxy formation at a level suitable for a graduate course in astrophysics. Conference proceedings are excluded, as are popular and semi-popular books.

*None* of the texts mentions the mass discrepancy–acceleration relation. Only two—Liddle and Loveday's *Oxford Companion to Cosmology* (2009), and Peter and Uzan's *Primordial Cosmology* (2013)—mention the existence of the universal acceleration scale $a_0$. Interestingly, the two textbooks that take a self-styled 'non-standard' view of cosmological theory—by Hoyle, Burbidge, and Narlikar (2005) and Capozziello and Faraoni (2011)—fail to mention either the acceleration scale or the mass discrepancy–acceleration relation.

## 7. Conventionalism *vs.* progress

Karl Popper believed that in order to maintain the falsifiability of a scientific theory, conventionalist stratagems should be avoided. It was argued above that some important elements of the standard cosmological model are conventionalist in the sense defined by Popper. At this point, it seems natural to depart from a strictly Popperian analysis and ask: Does it matter that the standard model is conventionalist? If so, why? And what are the implications, if any, for the progress of cosmology as a science?

Thomas Kuhn and Imre Lakatos expressed similar views on the role played by conventionalist stratagems in the progress of science. As summarized by Anthony O'Hear, Kuhn argued that

> during a period of what he [Kuhn] calls 'normal' science, the paradigm is protected from falsification. Counter-evidence to the paradigm, in the shape of falsified predictions, is treated as merely anomalous data, to be explained away in due course… Normal scientists will, in Popperian terminology, resort to conventionalist stratagems to deflect criticism from their paradigm. Sometimes these will produce new knowledge, as in the case of the discovery of Neptune. But sometimes they won't, as in the case of Mercury's orbit. (O'Hear, 1989, pp. 66-67.).

On this view, cold dark matter and dark energy might play a role similar to the role played by a postulated planet (Neptune) in explaining anomalies in Uranus's orbit. Or they might play a role like that of the mythical planet Vulcan, which was invoked (incorrectly) to explain the apsidal shift of Mercury's orbit.[32]

Like Kuhn, Lakatos recognized the tendency of scientists to invoke ad hoc hypotheses in order to protect the 'core' of a research program. Lakatos proposed a distinction between two sorts of incorporation, which he called progressive and degenerative:

> A research programme is said to be progressing as long as its theoretical growth anticipates its empirical growth, that is as long as it keeps predicting novel facts with some success ("progressive problemshift"); it is stagnating if its theoretical growth lags behind its empirical growth, that is as long as it gives only post-hoc explanations either of chance discoveries or of facts anticipated by, and discovered in, a rival programme ("degenerating problemshift") (Lakatos, 1971, pp. 104–105).

In a progressive research program, according to Lakatos, changes are driven by the program's inner logic, not by awkward empirical facts; when this ceases to be the case, the research program degenerates. Furthermore, when additions are made to a theory, Lakatos insisted that (at least some of) this "excess content" be corroborated by subsequent experiments or observations if the theory is to be considered progressive (Lakatos, 1970).

---

[31] Discussions of the observational evidence for an accelerating universe and its interpretation in terms of 'dark energy' began appearing in textbooks almost immediately after the papers in 1998 and 1999 that described the anomaly; examples are Rees (2001), Raine and Thomas (2001), and Ryden (2003).

[32] A more apt comparison here would be with phlogiston or luminiferous aether. Just as in the case of dark matter or dark energy, the assumed properties of these hypothesized substances were modified over time in response to new experimental data; see e.g. Partington and McKie (1981).

As with all theories of scientific progress, Lakatos' criteria can be criticized, in respect to their generality or their correspondence with the historical record. Here I take Lakatos' ideas "as read" and ask what they imply for the standard model of cosmology.

The current arc of cosmological theory clearly fails to meet the standards set by Lakatos for a progressive research program. Nothing in the pre-existing model (ca. 1970) pointed toward the need for dark matter or dark energy; the observations that motivated these hypotheses came as a complete surprise. Nor was the mass discrepancy–acceleration relation anticipated before it was established observationally.[33] Furthermore a great deal of current effort in the field—perhaps the bulk of the effort—is directed toward coping with the discrepancies. It has been quite a long time —perhaps not since the 1960s—that developments in the field were driven by its inner logic, rather than by unanticipated observational facts (Kroupa, 2012).

The evolution of the standard model has also clearly failed to satisfy Lakatos' criterion of "incorporation with corroborated excess content". The failure to directly detect the particles making up the dark matter—and hence to corroborate the dark matter hypothesis—was detailed above. Mordehai Milgrom (1989) notes, more generally, that the dark matter hypothesis

> has, so far, proven a great disappointment as a scientific hypothesis. To my knowledge, it has not given rise to a single unavoidable prediction; it has not produced any hitherto unexpected relations between galactic phenomena. The DM hypothesis simply states that dark matter is present in whatever quantities and space distribution is needed to explain away whichever mass discrepancy arises.

At the same time, Lakatos did not recognize any rigorous criterion for deciding, on the *short* term, whether changes or additions to a theoretical system will contribute to its long-term success: "The old rationalist dream of a mechanical, semi-mechanical or at least fast-acting method for showing up falsehood, unprovenness, meaningless rubbish or even non-rational choice has to be given up…It is very difficult to decide, especially if one does not demand progress at each single step, when a research programme has degenerated hopelessly" (Lakatos, 1978, p. 149).

## 8. Convergence

There is however one criterion that is widely viewed—by many scientists, and by at least some philosophers of science—as indicating that a theory is successful: the criterion of convergence. When a host of different experiments that presuppose a certain theory produce results that are consistent and compatible with the theory—for instance, by yielding a common value for a certain physical constant—there is a strong motivation to accept the theory as correct, or at least as an advance over pre-existing theories that did not have this characteristic. In the words of Hilary Putnam (1975), such agreement would seem "miraculous" unless the underlying theory were correct.

Jean Perrin made this argument in 1913, in favor of the atomic theory of matter. Perrin noted that Avogadro's number appears in equations that describe many different phenomena: Brownian motion, electrolysis, radioactive decay, the black-body spectrum etc.:

> Our wonder is aroused at the very remarkable agreement found between values [of Avogadro's number] derived from the consideration of such widely different phenomena. Seeing that not only is the same magnitude obtained by each method when the conditions under which it is applied are varied as much as possible, but that the numbers thus established also agree among themselves, without discrepancy, for all the methods employed, the real existence of the molecule is given probability bordering on certainty. (Perrin, 1913, pp. 215–216.).

Max Planck argued in a similar way that the agreement between different experimental determinations of the constant $h$ (Planck's constant) constituted strong support for the quantization of energy (Planck, 1922, pp. 496–500).

A similar argument is often made by cosmologists with regard to the standard cosmological model. The argument goes as follows:[34] The universe (on sufficiently large spatial scales, and as described in terms of Einstein's equations) can be parametrized in terms of a small set of numbers; at a minimum, six or seven are required.[35] These include the current mean densities of normal matter, dark matter, and dark energy; the current expansion rate; and the parameters that define the initial power spectrum of density perturbations. Now, there turns out to exist a choice for this set of numbers that is consistent with most of the observational tests that have been devised to date. This 'concordance' (the term favored by cosmologists) between different data sets is said to be compelling evidence for the correctness of the standard model, and by extension, for the existence of dark matter and dark energy. For instance, Olive et al. (2014) write: "The concordance model is now well established, and there seems little room left for any dramatic revision of this paradigm."

But the concordance argument made by Olive and others is substantially weaker than the convergence arguments made by Perrin and Planck, for a number of reasons.

**1**. In the instances discussed by Perrin (Avagadro's number) and Planck (Planck's constant), it was the agreement of the measured value of a *single* parameter, in *multiple* experiments, that lent credence to the reality of atoms and energy quantization respectively. But most of the parameters that define the concordance cosmological model have not been determined independently of each other. This is because—in most of the observational tests— there is substantial degeneracy between the best-fit values of two or more of the parameters. One example was mentioned above: in applying the 'Hubble diagram' test,[36] there is degeneracy between the parameters that define the matter density and the dark energy density: the larger the value assumed for one, the larger the value implied for the other. A similar degeneracy, but in the opposite sense, exists in tests based on angular fluctuations in the cosmic background radiation: in this case, the larger the assumed matter density, the smaller the implied dark energy density (Kowalski et al., 2008). A third example is the angular power spectrum of galaxies, which is predicted to evolve in a particular way as galaxies and their attendant dark matter cluster gravitationally; these data provide a useful constraint on the mean matter density but contain almost no useful information about the dark energy density (Eisenstein et al., 2005).[37] When cosmologists speak of 'concordance', they mean that it is possible to find a single *set* of parameters that provides an acceptable fit to the *conjunction* of observational data sets, and not that there is independent confirmation of the value of any single parameter. This is a fundamentally weaker sort of convergence than what Perrin and Planck were talking about,[38] and it is correspondingly less compelling in terms of justifying the correctness of the ΛCDM model.

---

[33] At least, not by researchers working within the standard paradigm. See note 29.

[34] Versions of this argument can be found in many of the texts listed in Table 1, e.g. Hawley and Holcomb (2005, p.428), Schneider (2015, p. 457), Heacox (2016, p. 175).

[35] One of the seven numbers is the mean density of radiation in the universe. This quantity is directly measurable with good precision and is usually considered to be known.

[36] Often referred to by cosmologists as the "supernova test" since most of the distant data points in the Hubble diagram are derived from observations of supernovae.

*2*. Some of the parameters that define the concordance model *can* be well constrained from the observations, in a way that is independent or nearly independent of the other parameters. But the values so obtained tend to be at variance with the concordance values. One example is Hubble's constant, $H_0$, which measures the current rate of cosmological expansion. Determination of $H_0$ using the classical redshift-magnitude test (the 'Hubble diagram') has consistently yielded higher values than the 'concordance' value; a recent study (Riess et al. 2016) finds $H_0 = 73.03 \pm 1.79$ km s$^{-1}$ Mpc$^{-1}$ compared with the concordance value of $67.3 \pm 0.7$ km s$^{-1}$ Mpc$^{-1}$, a 'three-sigma discrepancy'. A more striking example concerns the mean density of normal (non-dark) matter; cosmologists call this the 'baryon density'.[39] The density of baryonic matter is well constrained by the requirement that nucleosynthesis shortly after the Big Bang produce the observed abundances of the light elements:[40] deuterium, helium, lithium and their isotopes. But there is a factor of two discrepancy between the baryon density inferred in this way from the observed lithium abundance and the concordance value (this is the 'Lithium problem' mentioned in Section 2).[41]

*3*. Discussions of the concordance model never take into account the mass discrepancy–acceleration relation. In one sense, this is justified by the fact that the concordance model defines the structure and evolution of the universe only on the largest spatial scales. But as discussed above, the mass discrepancy–acceleration relation conflicts in an apparently irreconcilable way with the interpretation of galaxy rotation curves in terms of dark matter. And as documented above, this awkward fact is simply ignored.

An even more fundamental point can be made about the convergence/concordance argument (Losee, 2004). Perrin, by noting agreement between different determinations of Avogadro's number, was arguing in favor of theories that incorporate an atomic-molecular description of matter. He was not arguing for any *particular* atomic theory. In the same way, Planck's argument was for a progression to theories that incorporate energy quantization; he was not claiming that the agreement between different measurements of $h$ warranted the selection of any particular version of (what we would now call) quantum mechanics. These scientists were arguing, on the grounds of convergence of the value of a measured parameter, for transition from one *type* of theory to another. They were not arguing for the correctness of any single theory about atomic interactions or energy quantization.

In the same way: even if one stipulates that convergence, in the strong sense discussed by Perrin and Planck, could one day be established with regard to the cosmological parameters, this in itself would not mandate belief in any particular cosmological model. It would only imply that a transition is warranted: from models (such as the standard model ca. 1970) that contained nothing corresponding to dark matter or dark energy; to models that include one or more 'dark sectors.' Convergence, on its own, would provide no evidence for the existence of particle dark matter, since there is nothing in the concordance model, or the observations that motivated it, that requires the 'dark matter' component to be particulate: the only apparent requirement is that this component respond gravitationally, on large scales, like ordinary matter. And on smaller (galactic) scales, the dark sectors of the new model would need to reproduce the mass discrepancy–acceleration relation, something which the current model makes no attempt to do.

## 9. Discovery in cosmology

The expression "conventionalist stratagem" seems to imply *intent*: intent to maintain the integrity of a theory by circumventing a falsifying experiment. Nothing in Popper's definitions of conventionalist stratagems necessarily implies intent, however, and the arguments given above are likewise unaffected by the (unknowable) motivations of cosmologists. But it is natural to ask how cosmologists themselves describe the elements which here have been described as conventionalist.

In the textbooks listed in Table 1, it is easy to find statements like the following (emphasis added):

> *We know that the cosmological constant [Λ] exists* because of its gravitational effect. *This is the same as the situation for the CDM [cold dark matter]* and for that reason, the vacuum energy is sometimes called dark energy. [Lyth (2016), p. 53]
>
> Even more surprising than *the existence of dark matter* is *the discovery that about 70% of the Universe consists of something that today is called vacuum energy, or dark energy*, and that is closely related to the cosmological constant introduced by Albert Einstein. [Schneider (2015), p. 5]
>
> Before *the observational discovery of dark energy* in 1998, most people believed that the cosmological constant is exactly zero and tried to explain why it is so. [Matarrese et al. (2011), p. 343]
>
> *The discovery that almost three-quarters of the present cosmic energy density is to be ascribed to an almost uniform dark energy component* able to produce, via its negative isotropic pressure, the accelerated expansion of the Universe, represents the most severe crisis of contemporary physics. [Matarrese et al. (2011), p. xi].

These statements suggest that—rather than conceive of dark matter or dark energy as postulates invoked in response to falsifying observations—cosmologists interpret those same observations as tantamount to the *discovery* of dark matter or dark energy. The logic seems to be:

1. Newton's theory of gravity and motion is correct (in the weak-field regime appropriate to galaxies).
2. In the absence of unseen mass, Newton's laws imply that galaxy rotation curves must fall.
3. Galaxy rotation curves are observed to be asymptotically flat.

---

[37] In applying this and other tests, it is common practice simply to fix certain of the parameters based on theoretical preconceptions; for instance, the universe is often assumed to have zero curvature even when the data being confronted do not require it.

[38] Because, for instance, the number of cosmological parameters to be determined is comparable to the number of tests that constrain them. By contrast, Perrin listed sixteen experiments that yielded independent measurements of Avagadro's number.

[39] 'Baryonic matter' also includes leptons (electrons), mesons, and all the other standard-model particles.

[40] Only light elements undergo nuclear reactions at sufficiently high rates for their abundances to be changed significantly during this short interval of time.

[41] The baryon density inferred from the abundances of the other light nuclei (deuterium, helium) was originally consistent with the low value inferred from lithium—a fine example of convergence. But following publication of the concordance model around 2000, the baryon densities inferred from these other two elements have drifted upward, so that both values are now consistent with the concordance value and inconsistent with the lithium value. Results obtained from lithium have remained unchanged. The reasons for this are not entirely clear; confirmation bias has been suggested (McGaugh, 2015).

∴ There must be dark matter.

or:

1. Einstein's theory of gravity and motion is correct.
2. In the absence of a universal component with the properties of dark energy, Einstein's equations imply that the cosmological expansion rate must decrease over time.
3. The expansion rate is observed to increase over time.

∴ There must be dark energy.

The assumption that Newton's or Einstein's theories *must be* correct, even in the face of falsifying instances, recalls Popper's words: "Whenever the 'classical' system of the day is threatened by the results of new experiments which might be interpreted as falsifications according to my point of view, the system will appear unshaken to the conventionalist."

### Acknowledgements

Comments by the two anonymous referees were very helpful in clarifying my arguments. I also am indebted to L. Blanchet, B. Famaey, A. Fine, P. Kroupa, S. McGaugh, and M. Milgrom for comments and criticisms that improved the presentation. I owe a special debt to Evelyn Brister for her advice and encouragement throughout the course of this work. I thank Will Taylor for his help in using the DMTools software (dmtools.brown.edu), with which Fig. 1 was made. Stacy McGaugh kindly provided the data plotted here in Figure 2. This work was supported by the National Science Foundation [Grant no. AST 1211602] and by the National Aeronautics and Space Administration [Grant no. NNX13AG92G].